# Nano-capattery: Taming electron traffic for a 367% leap in biohydrogen surge through suppressing competing pathways in photo photo-fermentative system

Muhammad Shahzaib[1],

[1]*Collaborative Innovation Center of Biomass Energy, Henan Agricultural University,*

*Zhengzhou 450002*

Corresponding author: Muhammad Shahzaib

engr.shahzaib04899@yahoo.com

## Abstract

The electron flux diverts electrons from optimal hydrogen production pathways to competitive pathways, which overall reduces the efficiency of the photo fermentation hydrogen production (PFHP) system. For tackling electron flux and metabolic pathway regulation, a hybrid material (nano-capattery (NC)) was developed based on cobalt-iron-nitrogen doped biochar (Co-Fe-NBC). The NC possessed both the capacitor property (287.91 F/g) and battery-like charge storage 38.3 mC/g with the highest energy density of 159.95 mWh/g. These properties existed due to its $Fe^{2+}/Fe^{3+}$ and $Co^{2+}/Co^{3+}$ redox cycle ability, a highly porous surface (291.81 $m^2$/g BET surface area) caused by the defects ($A_D/A_G$ 3.13) and abundant oxygen vacancies (OVs) observed through electro paramagnetic resonance. During PFHP, there is an 85% reduction in propionic acid, a 65.3% record electron management efficiency, an improved 1.34 $NAD^+$/NADH ratio, along with a 87% increase in dehydrogenase activity, confirming the superior role of NC in efficient regulation of the metabolic pathway and electron flux management. These exceptional properties of biochar-based NC raised the cumulative hydrogen production from 151.03 mL (control) to 589.54 mL, which was an enormous increase of 367%.

## 1 Introduction

Photo-fermentative hydrogen production (PFHP) from lignocellulosic biomass is a highly promising clean energy carrier, notable for its high energy density (122 kJ/g) and zero carbon emissions upon combustion, with various characteristics such as wide range of waste utilization capacity, minimum operational cost and solar energy utilization [1-3]. However, its practical application is hampered by the various factors specially competing metabolic pathways which leads to the production of the by-products which hinder the production of biohydrogen production [4]. The electron management efficiency (EME) is the backbone of any biochemical process; similarly, high organic loading rates create a critical electron surplus in the PFHP system, where the electron supply from substrate oxidation exceeds the capacity of terminal electron acceptors [5-8]. This imbalance raises hydrogen partial pressure, disrupts interspecies electron transfer (IET) [9-11], and breaks the thermodynamic equilibrium necessary for volatile fatty acid (VFA) oxidation. As a result, metabolism shifts from biohydrogen production toward competitive pathways such as propionate and ethanol synthesis, leading to the build-up of inhibitory metabolites, which ultimately cause system failure [12-15], thus PFHP process highly depends on electron management, and its commercial success is limited by the lack of a robust electron management system [16].

Recently it has been shown that electron transferring within the fermentative system is the key factor which considerably affect the production of biohydrogen production [12], [13]. . However, such materials lack the integrated functionality to dynamically control electron flow, thus there is need of materials which not only simultaneously buffer transient electron overflow (a capacitor-like function) but also provide a sustained, redox-driven electron supply (a battery-like function) to maintain metabolic stability under fluctuating conditions. The emerging concept of a "nano-capattery," a hybrid material that combines the rapid charge/discharge of a capacitor with the enduring discharge of a battery, presents a promising solution to this question [17, 18]. A true capattery material can instantaneously absorb excess electrons during metabolic peaks, preventing the destructive feedback of high electron pressure, and then release them steadily during low-activity periods, ensuring a continuous electron supply to nitrogenase. This dual functionality is essential for stabilizing the internal redox environment and preventing the metabolic shifts that lead to system failure.

Among potential scaffolds for such a material, biochar stands out due to its large specific surface area, hierarchical porosity, low cost, and environmental friendliness [19, 20]. While pristine biochar contains fewer redox-active sites, poor electron-transfer capability, and lower catalytic and buffering capacity, which make it less desirable for such electron-management applications. However, doping pristine biochar with bimetallic nanoparticles (e.g., Fe/Co) can introduce abundant redox-active sites, increase catalytic efficiency and drastically enhance its electron transfer and management capabilities. The selection of Iron (Fe) and Cobalt (Co) is strategic; Fe is a ubiquitous and biocompatible element in microbial metabolic pathways, while Co is a key constituent of vitamin B12, a crucial cofactor in several enzymatic reactions, enhancing their biocompatibility and catalytic synergy [21-24]. A significant challenge with bimetallic doping, however, is metal aggregation and leaching, which reduces catalytic efficiency [25, 26]. This can be effectively mitigated by nitrogen doping, which creates stable Metal-Nx coordination sites (e.g., Fe-Nx, Co-Nx) that anchor the metals, prevent aggregation, and simultaneously enhance the material's conductivity and pseudo-capacitance [27-31]. Furthermore, the nitrogen functionalities, particularly pyridinic-N and pyrrolic-N, provide lone pairs of electrons that facilitate charge transfer and increase pseudo-capacitance, while graphitic-N improves overall electrical conductivity [29, 30].

Recently, researchers paid greater attention to functionalized/doped biochar adoption in different environmental remediation [32-34], bioenergy production [15, 35-38], and catalysis-based applications [39, 40]. Especially, the nitrogen-doped biochar, owing to its π-π stacked electrons [32] and nitrogen functionalities [41, 42], helped to facilitate microbial growth and improve the electron transfer in environmental remediation [42], bioenergy production, and catalysis-based systems [32, 41-43]. While the biochar doped with cobalt and iron shows extraordinary performance in environmental remediation [21], bioenergy production [44], and catalysis-based applications [45] due to abundant redox-active sites [46], tuned physiochemical and electrochemical properties , which facilitate enhanced electron transfer ability and catalytic efficiency [41, 47, 48]. Previous studies revealed that the combination of Co, Fe and nitrogen doping introduced unique properties in the biochar (exceptional regeneration capability of $Fe^{2+}/Fe^{3+}$ and $Co^{2+}/Co^{3+}$ redox active sites, highly active $Co_7Fe_3$ alloy, $Fe_2O_3$ and $Fe^0$ nanoparticles attached on the surface, providing unparalleled properties for electron transferring and metabolic pathway regulation [41, 49]. However, Co, Fe and nitrogen-doped biochar-based nano-capattery has never been developed or studied before, which has enormous potential in advanced techniques. Also, the biochar-based nano-capattery concept was never realized before in any biochemical process, specifically in the PFHP process. The incorporation of such material can successfully resolve the major bottlenecks of the PFHP process by regulating the metabolic pathway, improving electron transfer and management, along with providing a highly stable redox environment within the system.

In this study, we synthesize and evaluate a cobalt-iron-nitrogen doped biochar (Co-Fe-NBC) as a transformative nano-capattery designed to address the electron management bottleneck in PFHP. It was assumed that the enhanced conductivity, capacitance, catalytic efficiency, abundant redox active sites, and other physiochemical properties of Co-Fe-NBC-based NC will synergistically improve the electron management efficiency for electron flux and metabolic pathway regulation to improve biohydrogen production. The SEM, BET, XPS, FTIR, EPR and Raman spectroscopy were adopted to thoroughly characterize the Co-Fe-NBC material. Similarly, the CV, LSV, EIS, and Dunn's model analysis techniques were adopted to validate its capacitance-like properties (current contribution, specific capacitance, charge storage capacity, energy storage ability). Batch PFHP experiments were conducted to assess its impact on biohydrogen production (cumulative biohydrogen production (CBP), hydrogen production rate (HPR)), redox environment (oxidation-reduction potential (ORP), $NAD^+$/NADH ratio), and metabolic pathways (VFAs profile, dehydrogenase activity). This work establishes a new paradigm for biochar-based NC, transforming it from a passive electron shuttle into an active redox mediator that can dynamically interact with microbial metabolism, with broad implications for scalable bioenergy and microbial electrosynthesis technologies.

## 2 Materials and Methods

### 2.1 Synthesis of biochar-based NC

Corn straw samples were collected from a nearby agricultural field in Henan province, China. Raw samples were air-dried at room temperature before being pulverized with a mechanical herb grinder up to a 60-mesh sieve size (Ruian Baixin Machinery Factory, China). The samples' composition was determined using an F800 cellulose analyzer (Jinan Hanon Instrument Co., LTD., China). The primary constituents are 35.02% cellulose, 26.5% hemicellulose, and 7.03% lignin [50, 51].

For the preparation of pristine biochar (PBC), 50 g dried CS was put into a tube furnace and heated at a ramp rate of 10 °C/min to pyrolyze at 900 °C for 4 hr under vacuum conditions. After the pyrolysis, the furnace cooled down to room temperature naturally, and the prepared biochar was milled for 10 min with a pestle and mortar. After that, it was washed with ditilled water (DW) and ethanol to remove inorganic/organic impurities, dried for further utilization in a vacuum oven for 24 hr at 60 °C, and named PBC. The same method (washing/drying) was adopted for the synthesis of all types of biochar at the pre-pyrolysis, pyrolysis, and post-pyrolysis stages. After that, 50 g dried CS and 10% melamine (wt%) with 100 mL deionized water (DW) were stirred in a flask using a magnetic stirrer for 4 hr at 400 rpm. The obtained slurry was then put into a vacuum oven for 24 hr at 60 °C; after drying, the dried CS-melamine slurry was put into a crucible made of porcelain and pyrolyzed under the same conditions adopted for PBC, and this melamine-doped biochar is named nitrogen-doped biochar (NBC). To improve physio-electrochemical properties, 900 °C was selected because carbon-based material possessed higher stability with 17% higher capacitance than prepared at 700 °C [52].

Three flasks were prepared with varying amounts of metal precursors and followed the same process for drying, calcination, and washing of NBC. In 1st flask, 10g of NBC, 100 mL DW, and 10% (wt%) cobalt nitrate hexahydrate ($Co(NO_3)_2 \cdot 6H_2O$) were mixed. In comparison,

2nd flask contained 10g NBC, 100 mL DW, and 10% (wt%) ferric nitrate nonahydrate (Fe(NO$_3$)$_3$.9H$_2$O), and in 3rd flask, 10g NBC, 100 mL DW, 5% (wt%) cobalt nitrate hexahydrate (Co(NO$_3$)$_2$.6H$_2$O) along with 5% (wt%) ferric nitrate nonahydrate (Fe(NO$_3$)$_3$.9H$_2$O) were added. Prepared samples were named Co-NBC, Fe-NBC, and Co-Fe-NBC, respectively. After washing and drying, all five types of biochar were again milled with a pestle and mortar until uniformity and stored in glass tubes for further utilization.

## 2.2 Characterization of biochar-based NC

The shape and structures of all biochar-based NC were examined using a SEM (ZEISS Gemini SEM 300). X-ray diffraction (XRD) was adopted to investigate the crystallinity and graphitization of all biochar-based NC using a Bruker D8 Advance [53]. The BET specific surface area and pore size were determined using the Microtrac BELSORP-mini II apparatus [14]. X-ray photoelectron spectroscopy (XPS) was utilized to analyze the elemental chemical state of all biochar-based NC, employing a Thermo Scientific K-Alpha spectrometer with 300W Al-Kα radiation [54]. Raman spectroscopy was conducted using a LabRam HR Evolution (HORIBA) at a wavelength of 532 nm. The analysis employed a laser line in backscattering geometry at room temperature to identify the stretching and bending vibrational modes of the samples. A 50 times objective was utilized over the range of 200-1800 cm$^{-1}$ [12]. A Bruker Vertex 80V instrument was employed for Fourier Transform Infrared (FTIR) measurements [4].

## 2.3 Electrochemical analysis of biochar-based NC

Electrochemical characterization of PBC, NBC, Co-NBC, Fe-NBC, and Co-Fe-NBC was performed using a CHI760E workstation with a three-electrode system: Ag/AgCl reference, Pt counter, and working electrodes comprising fluorine-doped tin oxide (FTO) plates coated with biochar samples [53]. The electrolyte consisted of 0.1 M sodium sulfate (pH 7). FTO substrates (1 cm$^2$) underwent sequential cleaning with isopropyl alcohol, acetone, and distilled water before drying at 100 °C. For electrode fabrication, 20 mg of each biochar was dispersed in 40 mL acetone via 10-minute sonication, followed by drop-casting onto FTO. Prepared electrodes were oven-dried at 90 °C for 12 hours prior to testing. Cyclic voltammetry (CV) spanned -1 to 1 V at 10-50 mV/s scan rates to assess redox behavior, while electrochemical impedance spectroscopy (EIS) employed a 10 mV AC amplitude across 1 Hz-10 MHz.

Further CV analysis employed Dunn's theoretical model to deconvolute charge storage mechanisms, where current density I($v$) relates to scan rate $v$ as mentioned in Eq. (1) [17];

$$I(v) = k1(v) + k2(v)^{1/2} \qquad (1)$$

Here, I(*v*) indicates current density while k1(*v*) represents capacitive-controlled contributions and k2(v)$^{1/2}$ diffusion-controlled processes [55]. Constants k1 and k2 were determined by plotting I(v)/(v)$^{(1/2)}$ versus (v)^$^{(1/2)}$, deriving k1 from the slope and k2 from the intercept of the plot line. Specific capacitance (C), specific capacity (Qs), and energy density (ED) were subsequently calculated using established Eqs. (2-4):

$$C = \frac{\int IVdV}{2vm\Delta V} \qquad (2)$$

$$Specific\ Capacity\ (Qs) = \frac{1}{2}C(\Delta V)^2 \qquad (3)$$

$$Energy\ Density\ (ED) = \frac{C \times (\Delta V)}{2 \times (3600)} \qquad (4)$$

## 2.4 Feedstock preparation

The raw lignocellulosic biomass of corn stover (CS) was prepared by following the procedure described in a previous study [56]. Further information regarding the preparation of the growth medium is provided in the supplementary file.

## 2.5 Preparation of the inoculum

Photo fermentative inoculum consists of species capable of biohydrogen production, such as *Rhodopseudomonas palustris* [51]. Further information regarding the preparation of the growth medium is provided in the supplementary file.

## 2.6 Batch PFHP experiment

The standard PFHP reactor (control) without incorporation of any material was prepared by following the procedure described in our previous studies [4, 12-14, 51, 56]. This control reactor is assumed to be the baseline for comparison to all other biochar-based NC. However, five consecutive incremental concentrations of 10, 20, 30, 40, and 50 mg/L were incorporated to determine the effect of all biochar-based NC in PFHP [4, 12-14, 51]. Detailed information regarding the preparation of the growth medium is provided in the supplementary file.

## 2.7 Analytical methods

The intracellular concentrations of NADH in PNSB (*Rhodopseudomonas palustris*) were quantified utilizing an NAD$^+$/NADH Assay kit, employing a colorimetric WST-8 protocol (Beyotime Biotechnology, Haimen, China) [57, 58]. The analysis of dehydrogenase activity in the bacterial culture was conducted through the reduction of TTC (2,3,5-triphenyl tetrazolium chloride), as has been previously recorded [59]. Gas chromatography (GC-112N, INESA, China) was employed to quantify biohydrogen concentration and volatile fatty acid concentrations in the fermentation broth, utilizing HPLC (Agilent). The DNS methods were used to examine the fluctuation in reducing sugar (RS) throughout the batch PHFP. The pH was measured by an FE20 pH meter (Mettler Toledo, China) [51]. All the experiments were run in triplicate to ensure the accuracy and reliability of the results. Comprehensive details of all the analytical methods are available in the supplementary file

## 3  Results and Discussion

### 3.1  Morphological, structural and physicochemical characteristics of biochar-based NC

The morphology of the samples was analyzed using SEM, as shown in Fig. 1(a) and Fig. 1(b), which display the structures of PBC and Co-Fe-NBC, respectively. Fig. 1(a) indicates that before doping, the surface of PBC appeared relatively scaly, with a crystalline mesoporous structure and a visible irregular porous pattern. After doping, the Co-Fe-NBC surface was notably rougher, featuring both micro- and mesopores with clearly open channels, indicating a highly porous structure with a larger surface area. Additionally, it can be seen that many irregular small particles are attached to the surface in Fig. 1(b), most of which are likely Fe, and Co, ($Fe^0$, $Fe_2O_3$, $Co_2Fe_3$ Alloy) formed during high-temperature doping of metals in pyrolysis [60]. PBC mainly contains C (79.87%) and O (19.62%), with small amounts of naturally occurring S (0.43%) and Fe (0.08%). Meanwhile, the atomic compositions of C, N, O, Fe, and Co in Co-Fe-NBC were 52.71%, 1.93%, 35.58%, 4.72%, and 5.06%, respectively (Fig. S1). Elemental mapping of C, O, N, Fe, and Co in Co-Fe-NBC is shown in Fig. 1 (c, d, e, f, g), respectively. Numerous mesopores are present on PBC, while the surface of Co-Fe-NBC features both mesopores and micropores, indicating a high capacity for cell immobilization. The porous structure also provides a suitable environment for microbial colonization, thereby enhancing the PFHP process. Research shows that microbes tend to adhere better to rough surfaces, which increases the specific surface area for cell attachment [61]. Fig. 1(b) shows that Co-Fe-NBC has a more complex micro-porous structure, with a variety of sizes and more defects, which likely improves catalytic activity by providing additional active sites.

The surface area and porosity of the biochar-based NC were examined using nitrogen adsorption and desorption measurements, as shown in Fig. 1(h). It can be observed that PBC exhibited a type IV hysteresis loop containing mesopores, with a BET-SSA of 106.01 $m^2g^{-1}$, a pore volume of 0.0357 $cm^3/g$, 1.3185 nm pore diameter, respectively. The pore size distribution curves of NBC, Co-NBC, Fe-NBC, and Co-Fe-NBC indicated that their isotherms belonged to both types I and type IV hysteresis loops, indicating the presence of micropores and mesopores. Furthermore, the BET-SSA of NBC increased to 207.31 m$^2$g$^{-1}$, while Co-NBC, Fe-NBC, and Co-Fe-NBC reached 284.74, 288.56, and 291.81 m$^2$g$^{-1}$, respectively, after doping with N, Co, and Fe (Table 1). The highest BET-SSA was observed in Co-Fe-NBC, which was 69.12%, 40.76%, 2.48%, and 1.13% higher than PBC, NBC, Co-NBC, and Fe-NBC, respectively. Both single and dual metal doping significantly enhance BET-SSA compared to PBC and NBC due to their catalytic role during pyrolysis, which not only increases the BET-SSA and active sites for improved catalytic performance but also boosts total pore volume by increasing micropore and mesopore volumes. The BET-SSA of biochar greatly influences the surface reactivity of these carbon-based catalysts. [62]. The elevated specific surface area and nanoscale dimensions of biochar are advantageous for the PFHP, as this increased catalyst-electrolyte (fermentation broth) contact area enhances ion diffusion and exposure of a higher number of catalytically active sites. Additionally, the size of the pores plays a significant role; for instance, a mesoporous structure encourages microbial proliferation, while a micropore structure offers active sites that enhance catalytic activity [63]. Therefore, a suitable blend of micropores and mesopores would enhance the catalytic performance of biochar.

**Table 1: BET-SSA, BJH meso, and micropore analysis of all biochar-based NC**

| Sample | BET-SSA (m$^2$g$^{-1}$) | Mesopore Volume (cm$^3$/g) | Mesopore Diameter (nm) | Micropore Volume (cm$^3$/g) | Micropore diameter (nm) | Vmeso/ Vmicro |
|---|---|---|---|---|---|---|
| PBC | 106.01 | 0.036 | 1.319 | --- | --- | --- |
| NBC | 207.31 | 0.055 | 1.241 | 0.117 | 1.464 | 0.470 |
| Co-NBC | 284.74 | 0.108 | 2.628 | 0.158 | 1.465 | 0.684 |
| Fe-NBC | 288.56 | 0.111 | 2.594 | 0.164 | 1.496 | 0.677 |
| Co-Fe-NBC | 291.81 | 0.109 | 2.663 | 0.152 | 1.415 | 0.716 |

An increasing trend in mesopore and micropore volume from NBC < Co-NBC < Fe-NBC was observed, having values (0.0550, 0.117 cm$^3$/g), (0.108, 0.158 cm$^3$/g), and (0.111, 0.164 cm$^3$/g), respectively. However, Co-Fe-NBC achieved higher (0.109, 0.152 cm$^3$/g) mesopore and micropore volume than NBC and Co-NBC but lower than Fe-NBC. Although

Co-Fe-NBC possessed lower mesopore and micropore volume than Fe-NBC, it contains the highest BET-SSA and Vmeso/Vmicro ratio conducive to higher catalytic efficiency. Especially, Vmeso/Vmicro is a fundamental parameter to observe the electrochemical performance of biochar because the porous framework and elevated surface area contribute to improved contact between the biochar and electrolyte interface to minimize the diffusion resistance, thereby enhancing the capacitance and redox behavior of biochar. Consequently, an increased ratio of Vmeso/Vmicro enhances ion diffusion by these shorter pathways, promoting the penetration of the electrolyte (fermentation broth) into the pores and boosting the adsorption of ions on the biochar surface [64].

The XRD analysis was performed to investigate the crystallinity and graphitization of the synthesized Co-Fe-NBC-based NC. All samples showed a broad diffraction peak near 24° (2θ), related to the (002) plane of amorphous carbon (Fig. 1(i)). The PBC matched JCPDS 98-006-1781 (hexagonal structure), while NBC aligned with JCPDS 98-006-7245 (monoclinic structure). The Co-NBC displayed a prominent peak at 43.5°, indexed to metallic Co (JCPDS 98-024-0211, anorthic structure), supporting XPS findings of surface oxidation ($Co^{3+}$ at 780.71 eV) and nitrogen coordination ($Co^{2+}$-N at 785.1 eV). The Fe-NBC exhibited dual phases with peaks at 41.5° for $Fe_2O_3$ (JCPDS 98-009-9821, orthorhombic structure) and 44.4° for metallic $Fe^0$ (JCPDS 98-028-1494, anorthic structure), indicating partial carbothermal reduction. Most importantly, Co-Fe-NBC showed three clear peaks at 42.8° (110) and 65.5° (200), consistent with a face-centered cubic $Co_7Fe_3$ alloy (JCPDS 98-016-2711, orthorhombic structure). The relevant stick patterns of all biochar-based NC are presented in Fig. S14-S18.

Raman spectroscopy is an effective method for analyzing materials' crystalline and defective characteristics, as illustrated in Fig. 1(j). Characteristic peaks were observed at approximately 1356 and 1588 cm$^{-1}$, corresponding to the D and G bands. The D band signifies the signal of disordered carbon resulting from vacancies, functional groups, or heteroatom doping. In contrast, the G band arises from the in-plane vibration of the graphitic structure. Signals were deconvolved using Gaussian fitting for simplification, and the area under the peaks was calculated (Fig. S2). The ratio of area under the D-band to the area of the G-band can be used to analyze the nature of biochar surfaces. In addition, the area ratio of the D versus G bands (AD/AG) can speed up the electrophilic reaction to produce reactive species and improve the flow of π-electrons in the carbon matrix by conjugation [65]. The AD/AG values of PBC, NBC, Co-NBC, Fe-NBC, and Co-Fe-NBC were recorded as 2.71, 2.58, 3.10, 2.82, and 3.33, respectively, Fig. 1(j). Fe and Co doping induced distortion within the carbon network of NBC, generating structural defects evidenced by an increased AD/AG ratio and decreased graphitization [66]. This results in a more amorphous structure, which exhibits a higher SSA than graphitic carbon, as confirmed by BET data (Table 1). The elevated SSA exposes a greater number of reaction sites, facilitating catalyst-substrate contact. Furthermore, the lattice defects introduced by doping are beneficial for anchoring and dispersing metal atoms across the biochar surface [67].

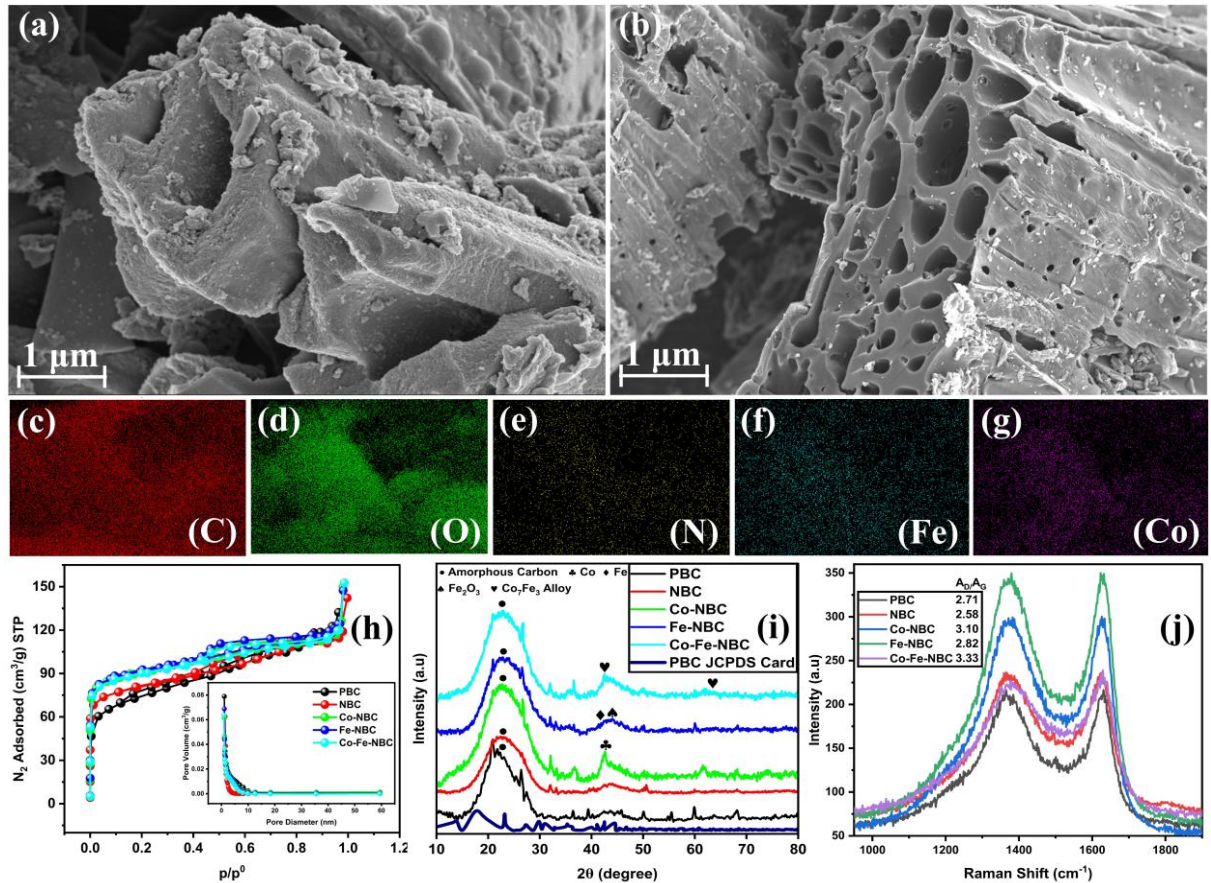

Fig. 1: Properties of biochar-based NC (a) SEM results of PBC, (b) SEM results of Co-Fe-NBC, (c, d, e, f, g) EDS elemental mapping of (C, O, N, Fe, Co), (h) BET (inset BJH), (i) XRD analysis, (j) Raman spectroscopy analysis

The XPS study characterized the surface elemental composition and oxidation states of PBC, NBC, Co-NBC, Fe-NBC, and Co-Fe-NBC based NC. The C1s spectrum (Fig. 2(a)) resolves into three components: C-C/C=C (~284.8 eV), C-O (~286.2 eV), and C-N (~287.6 eV). Corresponding N1s spectra (Fig. 2(b)) reveal four nitrogen configurations: pyridinic N (~398.1 eV), pyrrolic N (~399.4 eV), graphitic N (~400.8 eV), and chemisorbed nitrogen species (~402.5 eV) [41].

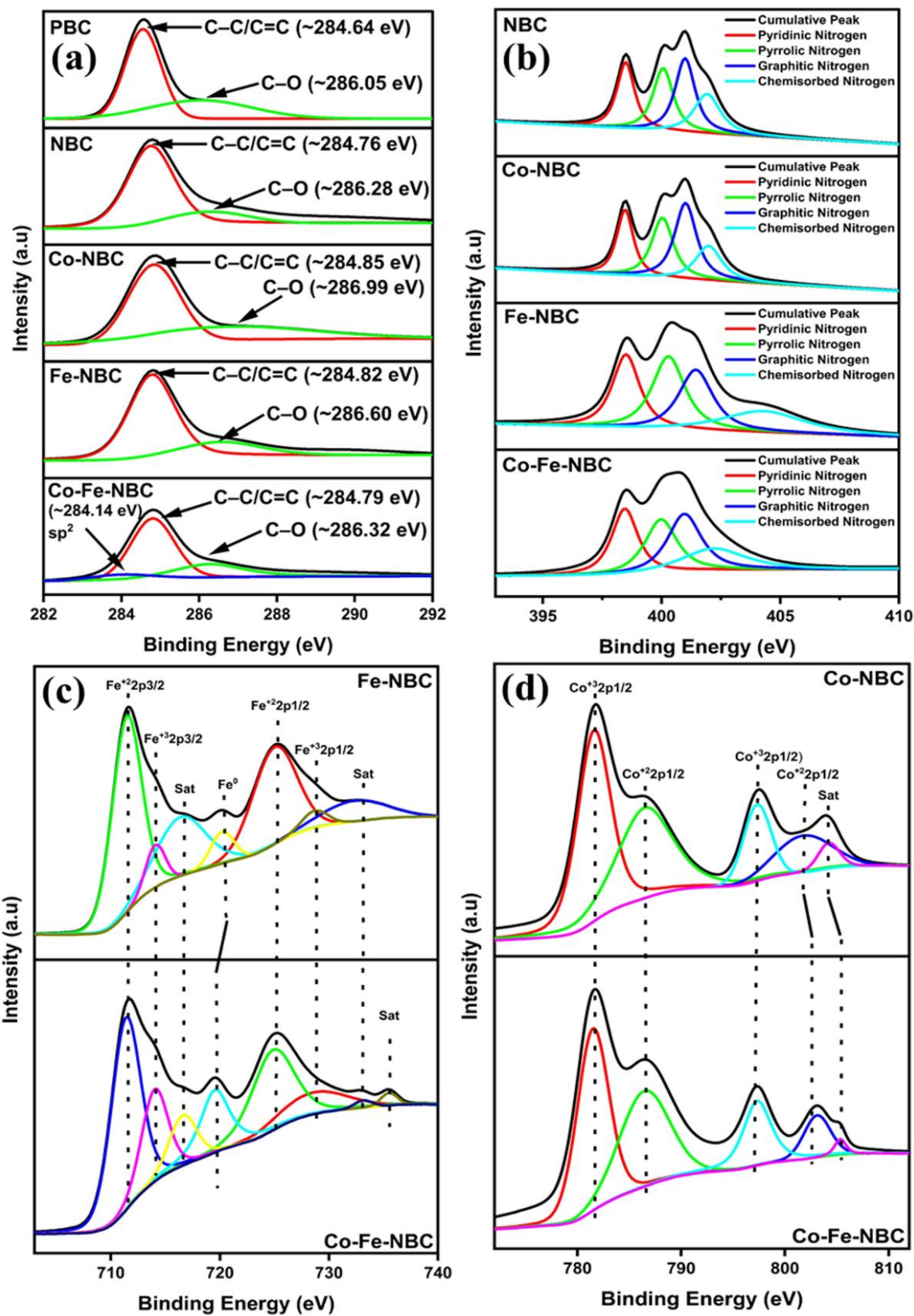

Fig. 2: XPS spectrum of (a) C1s, (b) N1s, (c) Fe2p, (d) Co2p

The Fe 2p spectrum (Fig. 2(c)) exhibits five distinct peaks: Fe(II) $2p_{3/2}$ (711.3 eV), Fe(III) $2p_{3/2}$ (713.8 eV), $Fe^o$ (720.4 eV), Fe(II) $2p_{1/2}$ (723.2 eV), and Fe(III) $2p_{1/2}$ (726.95 eV)

[68], with the 713.1 eV feature specifically indicating Fe(III)-N bonding [41]. High-resolution Co 2p spectra (Fig. 2(d)) display spin-orbit split $2p_{3/2}$ and $2p_{1/2}$ peaks. Co(II) states appear at 785.1 eV and 802.32 eV [69], while Co(III) signatures occur at 780.71 eV (Co $2p_{3/2}$) and 796.63 eV (Co $2p_{1/2}$) [48], with the 785.1 eV peak further confirming Co(III)-N coordination [41]. These results verify FeNx-CoNx moieties in Co-Fe-NBC. Coexistence of Fe(III) and Fe(II) indicates oxidation of $Fe^0$, while accelerated Fe(III)→Fe(II) conversion occurs via synergistic Co(II)/Co(III) redox cycling [70].

The EPR analysis enables characterization of surface oxygen vacancies (OVs) formation and nature in metal-functionalized biochar, where spectral signal intensity directly correlates with OVs concentration [71-73]. All three materials exhibit a singular Lorentzian peak at a g-value of 2.003 (Fig. 3(a)), indicative of oxygen-related local magnetic fields arising from the Zeeman interaction with unpaired electrons trapped at OVs sites [71, 73]. Comparative analysis reveals a pronounced hierarchy in EPR signal intensity: Co-Fe-NBC demonstrates the strongest response, followed sequentially by Fe-NBC and Co-NBC. This progression confirms that bimetallic doping induces the highest degree of oxygen atom removal, consequently amplifying both OV density and charge carrier mobility [26]. These EPR observations align with XPS and Raman spectroscopy data, collectively establishing that superior defect density primarily attributable to oxygen vacancies underpins the enhanced catalytic performance of Co-Fe-NBC [74]. Furthermore, the change in peak intensity and minor peak shifts also confirms the formation of OVs during redox cycling, as the change in the EPR analysis also confirms this phenomenon.

To simplify the FT-IR spectra, all biochar signals were deconvolved using Gaussian fitting (Fig. 3(b) & (Fig. S3)), and the changes in functional group positions and intensities were observed systematically. Standard peaks of O-H (alcohol/$H_2O$), C≡N (cyano/nitrile), C=O (carbonyl), C=O (ester/carboxylic acid/lactone groups), C=C/C=N (alkene/pyridinic N), C=C (aromatic ring or C=O in amide), C-N (Chemisorbed N), C-O (alcohols/ethers/esters), C-O or C-N (Chemisorbed N/amides/ethers), C-H (aromatic rings), C-H (alkanes/aromatic rings), C-N (Chemisorbed N), Fe-O, and Co-O were observed near 3715, 2399, 2313, 1773, 1711, 1650, 1575, 1474, 1220, 1100, 882, 819, 665, $cm^{-1}$ respectively. However, peaks related to nitrogen functionalities (NFs) (2399, 2313, 1575) and metal-O (665, 583, 447) were not present in PBC that were prominent in all other type of nitrogen and metal doped biochar, also a noticeable change in the position and intensity of peaks was observed which could be due to the stronger bond formation after hetero-materials doping [75]. Moreover, the highest intensity peaks at 1100 $cm^{-1}$ ascribed to C-N, which was beneficial to accelerate electron transfer in NFs containing biochar [76]. Also, the nitrogen-containing functional groups' presence increased from 13.71% to 52.43%, 34.64%, 40.97%, 41.19% in PBC, NBC, Fe-NBC, Co-NBC, and Fe-Co-NBC, respectively. Therefore, most of the functionalities in PBC were oxygen-based because only a limited amount of naturally present nitrogen was available to generate NFs. At the same time, in all doped biochar preparations, melamine was used as a doping source, which enhanced the NFs' presence and decreased the oxygen-containing functional groups. It is worth mentioning that NFs have a higher ability to induce redox reactions and improve the wettability of carbon materials for the electrolyte solution than oxygen-containing carboxyl (COOH) and epoxyl (COC) groups [77]. Thus, metal and NFs generate new redox sites for higher catalytic and enhanced electrochemical properties. The presence of NFs in NBC was also confirmed through the negative zeta potential of NBC, which arises due to the abundance of NFs on the surface. Similarly, metal-containing functional groups in Co-NBC, Fe-NBC and Co-Fe-NBC were also confirmed through positive values of zeta potential that could be attributed to the presence of metal-containing functional groups.

The point of zero charge (PZC) values indicate the pH at which each biochar has a neutral surface charge, which provides insights into the surface chemistry and potential interactions of the biochar particles in different pH environments. PZC is a fundamental parameter of materials in accessing their redox behavior, stability, and interaction with other substances in the suspension because at the PZC surface shows no charge, but at pH below the PZC level surface is positively charged, attracting electrons while above the PZC level surface is negatively charged having attraction for positively charged species to transfer electrons. This property can create a path for the speedy transfer of electrons from the oxidation center to the reduction center. Therefore, PZC was calculated for PBC, NBC, Co-NBC, Fe-NBC, and Co-Fe-NBC, which were 6.37, 6.46, 7.65, 6.83, and 5.89, respectively, Fig. 3(c). As can be seen from Fig. 1(c), Co-Fe-NBC shows the lowest $\Delta$pH and highest $\Delta$pH below and above the PZC level, which is quite the opposite to all other biochars, indicating its higher ability to maintain a redox potential in varying pH, which is crucial in the PFHP process. Also, the lowest PZC value of Co-Fe-NBC from all other biochar indicates its higher ability to attract and repel electrons in an acidic environment, which is a pivotal characteristic of a highly efficient catalyst in PFHP, as most of the PFHP process remains in acidic conditions.

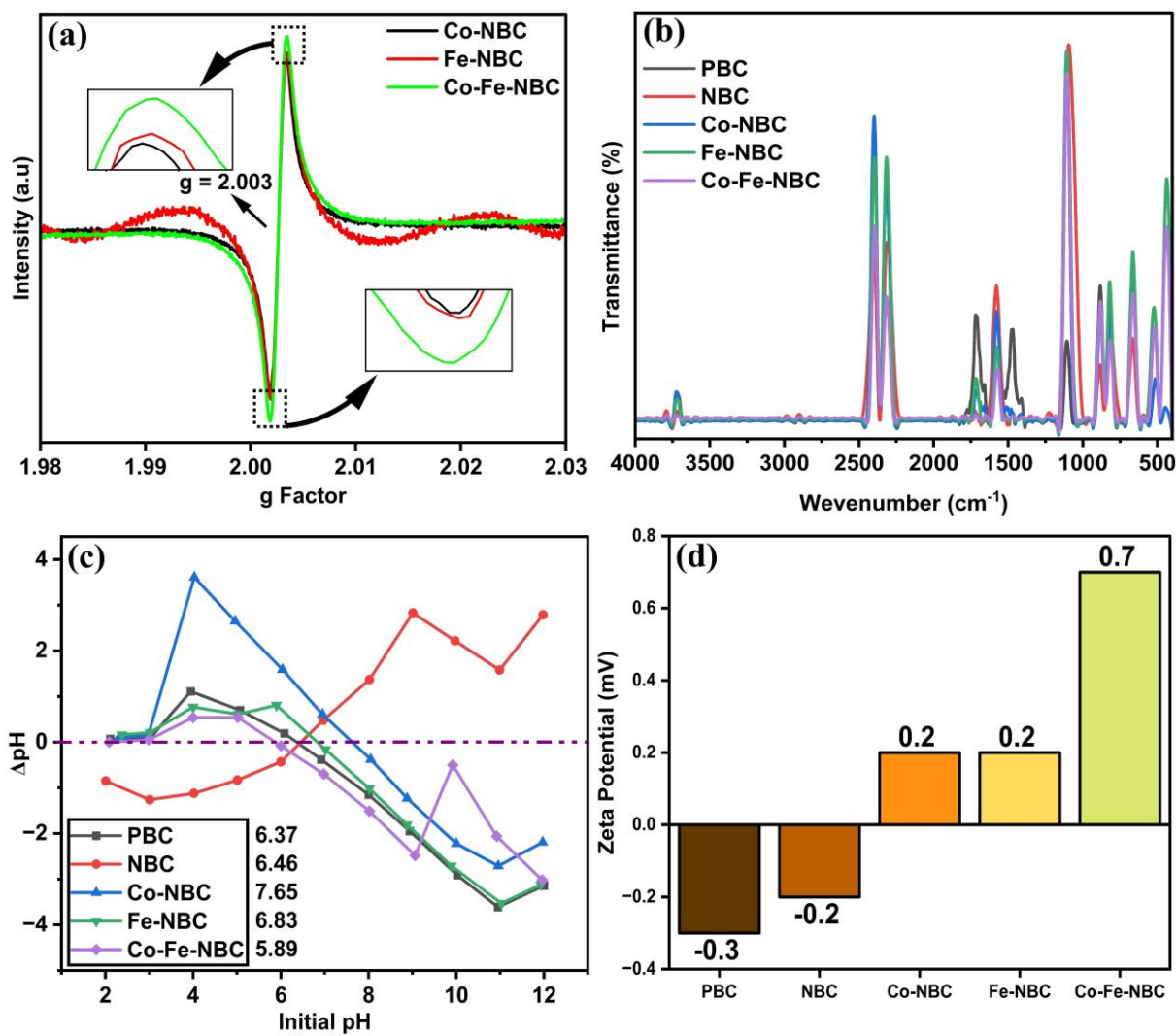

Fig. 3: Properties of biochar-based NC (a) EPR analysis, (b) FTIR analysis, (c) Point of zero charge study, (d) Zeta potential

A zeta potential study was conducted to assess the stability and agglomeration behavior of all biochar in colloidal dispersions, with the relevant results shown in Fig. 3(d). PBC exhibits a zeta potential of -0.3 mV, the lowest among all biochar samples, which correlates with a higher level of agglomeration; about 74% of the particles fall within a narrow size range of 0-0.5 μm. The negative zeta potential value could be attributed to the abundance of oxygen-containing functional groups on the surface. Although NBC shows a slightly higher zeta potential of -0.2 mV than PBC, it demonstrates a greater level of agglomeration, with 99.1% of particles confined within the 1.25-1.5 μm range. In comparison to PBC and NBC, positive zeta potential values of 0.2 mV, 0.2 mV, and 0.7 mV were observed for Co-NBC, Fe-NBC, and Co-Fe-NBC, respectively. However, both Co-NBC and Fe-NBC exhibit higher levels of agglomeration, with 92.24% and 97.77% of particles in the narrow size ranges of 1.5-1.75 μm. Compared to all other biochar samples, Co-Fe-NBC not only achieved the highest zeta potential but also showed significantly less agglomeration, with particles more evenly distributed within the broader range of 0-4.75 μm. This suggests that the combination of cobalt, iron, and nitrogen doping synergistically enhances the surface charge and stability of the biochar particles, likely due to improved electron transfer, increased active site availability, and better metal dispersion. Overall, Co-Fe-NBC demonstrated superior stability of the biochar particles in suspension compared to all other samples. These findings support recent research, suggesting that substantial improvements in biochar particles' stability and catalytic activity can be attained through surface chemistry and porosity modifications [78].

### 3.2 Electrochemical properties of biochar-based NC

The highest specific capacitance of PBC, NBC, Co-NBC, Fe-NBC, and Co-Fe-NBC was observed at 10 mV/Sec, which was 120.39 F/g, 160.76 F/g, 202.32 F/g, 217.80 F/g, and 287.91 F/g, respectively, Fig. 4(a). However, Co-Fe-NBC shows the highest specific capacitance of 287.91 F/g, 263.46 F/g, 246.94 F/g, 212.32 F/g, 183.29 F/g, and 132.19 F/g at different scan rates of 10, 20, 30, 40, and 50 mV/Sec, respectively (Fig. (a)). The highest capacitance of Co-Fe-NBC at 10 mV/Sec was 139.15%, 79.10%, 32.19%, and 42.30% higher than PBC, NBC, Co-NBC, and Fe-NBC, respectively. Furthermore, at a higher scan rate of 50 mV/Sec, Co-Fe-NBC shows the highest specific capacitance compared to all biochar. However, PBC shows the lowest specific capacitance from all types of biochar at all scan rates. This highlights that Co-Fe-NBC not only shows superior performance in a lower potential window but can also be effective when a significant change in voltage happens. Co-Fe-NBC is dominated by surface functional groups, NFs, and doping-based redox sites, which are conducive to the faster direct electron transfer (DIET) process in redox behavior [79]. Biochar donates electrons and creates a favorable environment for the growth of microorganisms. It can also provide adsorption sites for electron acceptors, directly correlating with improved catalytic activity.

Furthermore, the CV curve shows an anodic (oxidation) peak at 300 mV and cathodic (reduction) peaks at -120 mV and -700 mV on all scan rates. However, these peaks broaden at higher scan rates, which shows an increase in redox reaction when high-density current passes through the system. The appearance of these peaks confirms the occurrence of faradic reactions at the electrode surface and pseudo-capacitance behavior with a three-step redox process, along with charge transferring kinetics of all biochar [80]. This result also affirms the high catalytic activity of the electrode material and the reversible process for the charge storage. The CV curves maintain a consistent shape even at a scan rate of 50 mV/s, indicating a satisfactory current response of the Co-Fe-NBC at elevated scan rates, along with a high rate capability [81]. At lower scan rate of 10 mV/s, PBC, NBC, Co-NBC, Fe-NBC, and Co-Fe-NBC show a maximum specific capacity of 240.77 C/g, 321.51 C/g, 435.59 C/g, 404.64 C/g, and 575.81 C/g, respectively Fig. 4(b).

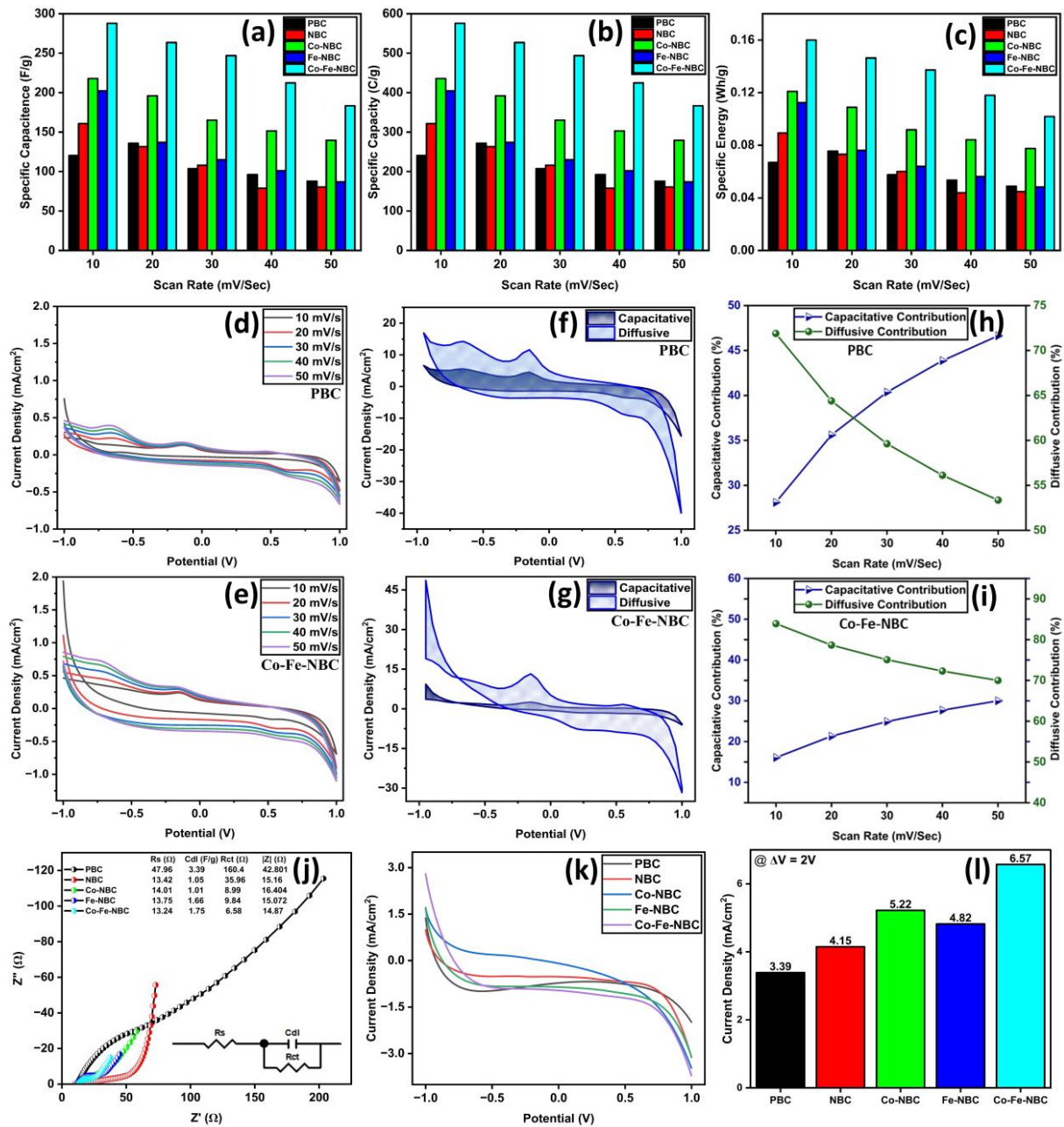

Fig. 4: Electrochemical properties of all NC (a) Specific Capacitance, (b) specific Capacity, (c) Specific Energy, (d) CV of PBC, (e) Current Distribution of PBC, (f) Current distribution w.r.t Scan Rate of PBC, (g) CV of Co-Fe-NBC, (h) Current Distribution of Co-Fe-NBC, (i) Current distribution w.r.t Scan Rate of Co-Fe-NBC, (j) EIS study, (k) LSV study, (l) Current density

A similar trend was observed in PBC, NBC, Co-NBC, Fe-NBC, and Co-Fe-NBC, which show a maximum specific energy density of 66.88 mWh/g, 89.31 mWh/g, 121 mWh/g, 112.4 mWh/g, and 159.95 mWh/g, respectively, at a lower scan rate of 10 mV/Sec Fig. 4(c). Both Specific capacity and specific energy density were gradually decreased with the increase in scan rate from 10 to 50 mV/Sec as depicted in Fig. 4(b, c), it could be justified through the current contribution which show that with the increase in scan rate capacitive contribution also increased because there is lesser time to store charges within the lattice so instead of storage flow of charges maintained through non faradic capacitive current. However, at a scan rate of 10 mV/Sec, Co-Fe-NBC shows a 139.15%, 79.09%, 32.19%, and 42.30% remarkable increase in specific capacity. In contrast, 139.15%, 79.09%, 32.19%, and 42.30% increases in specific energy density were recorded in comparison to PBC, NBC, Co-NBC, and Fe-NBC, respectively.

As illustrated in Fig. 4(d, e) and Fig. S5(a, b, c), the CV curves are not in a perfect rectangular shape, which suggests that the charge storage mechanism is not purely double-layer capacitance (DLC). Redox peaks or other features that deviate from a rectangular shape also indicate contributions from pseudo-capacitance. The curves show some degree of hysteresis, which is the difference between the forward (charging) and reverse (discharging) directions, and hysteresis is often associated with pseudo-capacitive behavior. Ideally, the capacitance should be relatively independent of the scan rate for a purely capacitive material. However, deviations can indicate the presence of pseudo-capacitance or other kinetic limitations. Therefore, the CV plot suggested that the material exhibits both double-layer and pseudo-capacitance. In addition, the relatively large CV curve integral area suggests that the Co-Fe-NBC has a higher specific capacitance [82]. This may be expected in the change in the oxidation state of iron and cobalt species as $Fe^{3+}/Fe^{2+}$, $Co^{2+}/Co^{3+}$ redox on the biochar surface, according to the result obtained in the XPS analysis (Fig. 2(c, d)) [83]. These redox species dominated the redox behavior of Co-Fe-NBC which were absent in all other biochar, which illustrates the decreasing trend in capacitance with an increase in scan rate from 10-50 mV/Sec owing to the rapid motion of charges under the high electric field.

The Fig. 4(f, g) and Fig. S5(d, e, f) illustrates the capacitive and diffusive current contributions derived from cyclic voltammetry at scan rates of 10, 20, 30, 40, and 50 mV/s, with the capacitive portion demarcated by light blue shading. Fig. 4(f) specifically quantifies the capacitive and diffusive percentages for PBC across these scan rates. At 10 mV/s, capacitive contributions were minimal: 28.10% for PBC, 19.73% for NBC, 6.05% for Co-NBC, 4.22% for Fe-NBC, and 16.08% for Co-Fe-NBC. Conversely, diffusive contributions dominated at 71.90%, 80.27%, 93.95%, 95.78%, and 83.92%, respectively. A significant shift occurred at 50 mV/s, where capacitive contributions increased to 46.64% (PBC), 35.47% (NBC), 12.58% (Co-NBC), 8.97% (Fe-NBC), and 20.99% (Co-Fe-NBC), while diffusive contributions decreased to 53.36%, 64.53%, 87.42%, 91.03%, and 70.01%, as shown in (Fig. 4(h, i) and Fig. S5(g, h, i). This inverse relationship shows that higher scan rates boost capacitive effects while reducing diffusive contributions. Consequently, the materials exhibit a behavioral transition from battery-like to capacitor-like characteristics with increasing scan rate. This occurs because lower scan rates permit sufficient time for complete redox reactions, favoring diffusion-controlled (battery-type) processes. In contrast, higher scan rates restrict charge adsorption times at the electrode/electrolyte interface, promoting kinetically controlled capacitive responses. These results also strengthen our claim about the biochar-based NC material, which is the most appropriate material to use as a catalyst in PFHP, as it can transfer high-energy electrons more efficiently due to the faradic-based redox reaction and also store surplus electrons and release them consistently due to battery-like ability for establishing a stable redox environment. These results are perfectly backed by Vmeso/Vmicro (Table 1), which shows that Co-Fe-NBC has the highest value of Vmeso/Vmicro, followed by the Co-NBC, Fe-NBC, and NBC, which is perfectly in order of specific capacitance.

To calculate the redox capacity of all types of biochar at varying scan rates, CV curves were convolved and differentiated using the built-in software of the CHI 760E electrochemical workstation. The highest total redox capacity of 15.8 mC/g, 22.8 mC/g, 27.7 mC/g, 27.0 mC/g, and 38.3 mC/g was recorded at a scan rate of 40, 20, 10, 10, and 10 for PBC, NBC, Co-NBC, Fe-NBC, and Co-Fe-NBC, respectively. Surprisingly, Co-Fe-NBC shows the highest TRC at 10 mV/Sec, which was 142.41%, 67.98%, 38.27%, and 41.85% higher than PBC, NBC, Co-NBC, and Fe-NBC, respectively. Similarly, PBC, NBC, Co-NBC, Fe-NBC, and Co-Fe-NBC show the highest oxidation capacity of 2.27 mC/g, 4.53 mC/g, 3.05 mC/g, 5.29 mC/g, and 4.94 mC/g, along with the highest reduction capacity of 11.90 mC/g, 18.26 mC/g, 27.65 mC/g, 11.90 mC/g, and 38.31 mC/g, respectively. The highest reduction capacity of Co-Fe-NBC was 221.85%, 109.79%, 38.53%, and 221.85% higher than PBC, NBC, Co-NBC, and Fe-NBC, respectively. The highest oxidation capacity from PBC, NBC, and Co-NBC increased by 117.99%, 9.14%, and 61.89%, respectively, but a slight decrease of -6.48% was observed when compared with Fe-NBC. Furthermore, this redox capacity was converted to calculate the highest number of electrons transferred to check the potential of all biochar in electron-mediated catalysis. As expected, Co-Fe-NBC shows the highest 13.24 µmole/g electron transferred, while PBC, NBC, Co-NBC, and Fe-NBC show the highest 5.45 µmole/g, 7.87 μmole/g, 9.32 μmole/g, and 9.55 μmole/g electron transferring ability in electron-mediated redox catalysis. The above results indicated that Co-Fe-NBC is more conducive to promoting the reactive species because it has a higher electron-exchange capacity [79].

An EIS analysis examines the electron transport phenomena and ion diffusion kinetics at the electrode/electrolyte interface, simulating the interaction between the catalyst and the fermentation broth [84]. The diameter of the semicircle associated with all biochar is directly proportional to its charge-transfer resistance, and the arc radius is analogous to the efficacy of electron transfer. It is well established that a smaller radius, coupled with lower electron transfer resistance, results in an improved electron transfer rate. As the arc radius of Co-Fe-NBC was smaller than that of all biochar, it was confirmed that Co-Fe-NBC has a faster electron transfer that resulted in its high photocatalytic performance (Fig. 4(j)) [85]. Further, the EIS of the PBC, NBC, Co-NBC, Fe-NBC, and Co-Fe-NBC was conducted at a range of 100-105 Hz. The recorded data of the Nyquist plot was processed in Zview software to fit an equivalent Randol's circuit for calculating important parameters like equivalent series resistance (Rs), charge transferring resistance (Rct), double layer capacitance (Cdl) and effective resistance |Z| (inset Fig. 4(j)). The Rs value of PBC, NBC, Co-NBC, Fe-NBC, and Co-Fe-NBC were recorded as 47.96 Ω, 13.42 Ω, 14.01 Ω, 13.75 Ω, and 13.24 Ω, respectively, which corresponds to the contact resistance of the current collector and the ohmic resistance of the interface. While the Rct across the interface of electrolyte and electrode of PBC, NBC, Co-NBC, Fe-NBC, and Co-Fe-NBC was recorded as 160.40 Ω, 35.96 Ω, 9.00 Ω, 9.84 Ω, and 6.58 Ω, respectively. |Z| is the effective resistance depicting the combined effect of imaginary resistance Z' and real resistance Z", which was recorded as 42.8 Ω, 15.2 Ω, 16.4 Ω, 15.1 Ω, and 14.9 Ω for PBC, NBC, Co-NBC, Fe-NBC, and Co-Fe-NBC, respectively. PBC shows the highest Cdl with 3.39 F/g, while NBC, Co-NBC, Fe-NBC, and Co-Fe-NBC show a lower Cdl of 1.05 F/g, 1.01F/g, 1.66 F/g, and 1.75F/g, respectively. However, here in this system, the Cdl contribution is insignificant, and many other parameters are more impactful in the PFHP process for electron management. From the above EIS characterization, all parameters suggest that the Co-Fe-NBC is the most efficient material with the lowest resistance, higher electron transfer capability, and enhanced catalytic efficiency, which is crucial for a material that can be used in enhanced photo fermentation biohydrogen production.

LSV of all biochar was performed to explore the electron transfer process and current response to correlate with electrical conductivity, Fig. 4(k, l). For PBC, the current densities of these biochar electrodes were minimal, being in the range of (1.4 to -1.99) mA/cm$^2$. The possible reason was that they had no transition metal or NFs available. Compared to the PBC, slightly higher current density from (1.01 to -3.13) mA/cm$^2$ was observed, which could be positively correlated with NFs (pyrrolic + graphitic) presence, which had a 49.25% of total nitrogen content. Because pyrrolic and graphitic NFs directly enhance the electrical conductivity in nitrogen-doped biochar [29]. In addition, graphitic nitrogen played a key role in the carbon materials, which was connected to the benzene ring on graphite to enhance the conductivity [86]. The current densities of Co-NBC and Fe-NBC electrodes were improved to (1.73 to -3.49) mA/cm$^2$ and (1.69 to -3.14) mA/cm$^2$, respectively. Compared with PBC and NBC, the current densities of Co-NBC and Fe-NBC were significantly increased due to the combined effect of transition metals and NFs' presence. Although Co-NBC has a higher content of NFs (60.79%), iron is more conductive than cobalt; therefore, Fe-NBC shows higher current density with slightly lower NFs (49.07%). However, Co-Fe-NBC shows the highest current density from (2.83 to -3.74) mA/cm$^2$ compared to all other biochars due to the combined effect of iron/cobalt and NFs. LSV results are correlating with the EIS study, which showed Co-Fe-NBC has the lowest Rs, Rct, |Z|, and a strong peak current due to Co and Fe loading, indicating that the introduction of Co and Fe dramatically accelerated the electron transfer on the Co-Fe-NBC surface and reactivity [87].

### 3.3 Influence of biochar-based NC on electron management in biohydrogen production and fermentation broth

Results of optimized dose of all biochar-based NC are illustrated in Fig. 5(a-k), which shows cumulative biohydrogen production (CBP), hydrogen production rate (HPR), hydrogen content (HC), reducing sugars (RS), pH, oxidation-reduction potential (ORP), acetic acid (A), butyric acid (BA), propionic acid (PA), and the current study compares with other published studies. The control gave 151.03 mL CBP, while optimized doses of PBC (40 mg/L), NBC (40 mg/L), Co-NBC (30 mg/L), Fe-NBC (30 mg/L), and Co-Fe-NBC (20 mg/L) yielded the highest CBP of 232.97 mL, 359.01 mL, 456.04 mL, 530.89 mL and 589.54 mL, respectively (Fig. 5(a)). Co-Fe-NBC outperformed all groups, with an astonishing increase of 290% and 153% compared to the control. Additionally, PBC was observed to show higher superiority due to its efficient electron management and metabolic activity. Along with CBP, Co-Fe-NBC also outperformed all other groups in HPR, with the highest value of 34.17 mL/h (Fig. 5(b)), which was 568.32%, 420.74%, 86.63%, 100.06%, and 116.36% higher than the control, PBC, NBC, Co-NBC, and Fe-NBC, respectively.

The effect of all biochar-based NC on RS generation is depicted in Fig. 5(c), which shows the concentration of RS with optimized dose, Co-Fe-NBC indicated its superior role in higher light capturing and facilitation of enzymes and PNSB for efficient degradation of biomass into sugars [88]. The RS is a building block in PFHP, as microbes consume the RS and convert it into VFAs, which are subsequently converted into hydrogen. Therefore, a higher yield of RS results in higher PFHP efficiency [88, 89]. The RS results show that the control accumulated the lowest concentration of RS, 1.83 g/L, but all biochar-based NC differentially modulated hydrolysis, which may be due to pore-driven enzyme immobilization, surface chemistry effects, and enhanced regulation of electron flux control. The highest RS concentration of 2.25 g/L, 6.51 g/L, 6.61 g/L, 7.7 g/L, and 10.42 g/L was observed in PBC, NBC, Co-NBC, Fe-NBC, and Co-Fe-NBC, respectively. All biochar-based NC outperformed control in RS generation, but Co-Fe-NBC showed extraordinary performance over its counterparts and gave 363%, 60%, 58% and 35% RS yield in comparison to PBC, NBC, Co-NBC, and Fe-NBC, respectively. These results indicate that Co-Fe-NBC exhibits higher enzymatic efficiency and superiority in promoting the growth of hydrolytic bacteria, as a higher yield of RS is directly attributed to the presence of hydrolytic bacteria [90].

Based on the produced metabolite concentrations in the PFHP experiment, it was observed that the PFHP system employed three different types of fermentation pathways, namely propionate-type (AA+PA), butyrate-type (AA+BA) [91], and ethanol-type fermentation (AA+BE) [92]. The highest accumulation of 5.25 g/L (AA+BA) was observed in the Co-Fe-NBC group, while the control, PBC, NBC, Co-NBC, and Fe-NBC, correspondingly produced the lower amounts of 0.92 g/L, 1.15 g/L, 3.28 g/L, 3.33 g/L, and 3.88 g/L (Fig. 5(d-e)). Surprisingly, the Co-Fe-NBC group produced the lowest concentration of 0.31 g/L PA, which was 85.07%, 71.64%, 78.24%, 79.14%, and 73.93% lower than the control, PBC, NBC, Co-NBC, and Fe-NBC, respectively (Fig. 5(f)). Higher AA+BA with the lowest PA generation confirmed the superior electron management capability of Co-Fe-NBC in handling the electron flux generated during the oxidation of RS and regulating metabolites to the most desired butyric acid pathway [14, 50]. At the same time, all other groups failed to show such electron management during electron flux and partially adopted the propionic acid pathway, which ultimately produced the lower yield of PFHP [4].

The adaptation of metabolic pathways is a complex process, as even slight changes in pH, ORP, and other factors can significantly influence it and lead to the transition to different types of metabolic pathways [4, 93]. During acidogenesis, Co-Fe-NBC not only regulated the metabolic pathway but also showed higher pH buffering capacity with minimum pH sustained above 5.7, as shown in Fig. 5(g). At the same time, all other groups failed to maintain 5.7 and dropped below this crucial limit, as this pH range favors the higher hydrogen-producing metabolic pathways while suppressing competing pathways [94]. Similarly, ORP results also correlate with our claim of capattery-like behavior of Co-Fe-NBC, as it achieved the lowest ORP of -512 mV, maintaining a higher potential difference within the environment to accelerate electron movement and yielding a higher PFHP [14, 50]. However, all other groups achieved a higher ORP value compared to the Co-Fe-NBC Fig. 5(h). A comparison of all biochar-based NC with one another, along with other previous studies, is illustrated in Fig. 5(i). These studies used different nano-photocatalysts to enhance PFHP yield by increasing surface area [14], electron transfer [13], OVs creation [12], decreasing electron hole pair recombination rate [12, 51, 53], but Co-Fe-NBC-based NC surpasses all these nano-photocatalysts in PFHP yield due to its higher EME, suggesting that electron management has a higher impact than other fundamental properties. For an in-depth study and better understanding, the detailed results of all biochar-based NC with all doses from 0 mg/L to 50 mg/L are illustrated in the supplementary file, in which Fig. S6, Fig. S7, Fig. S8, Fig. S9, Fig. S10, Fig. S11, Fig. S12 and Fig. S13 showing corresponding results of CBP, OPR, pH, RS, AA, BA, PA and (BE), respectively.

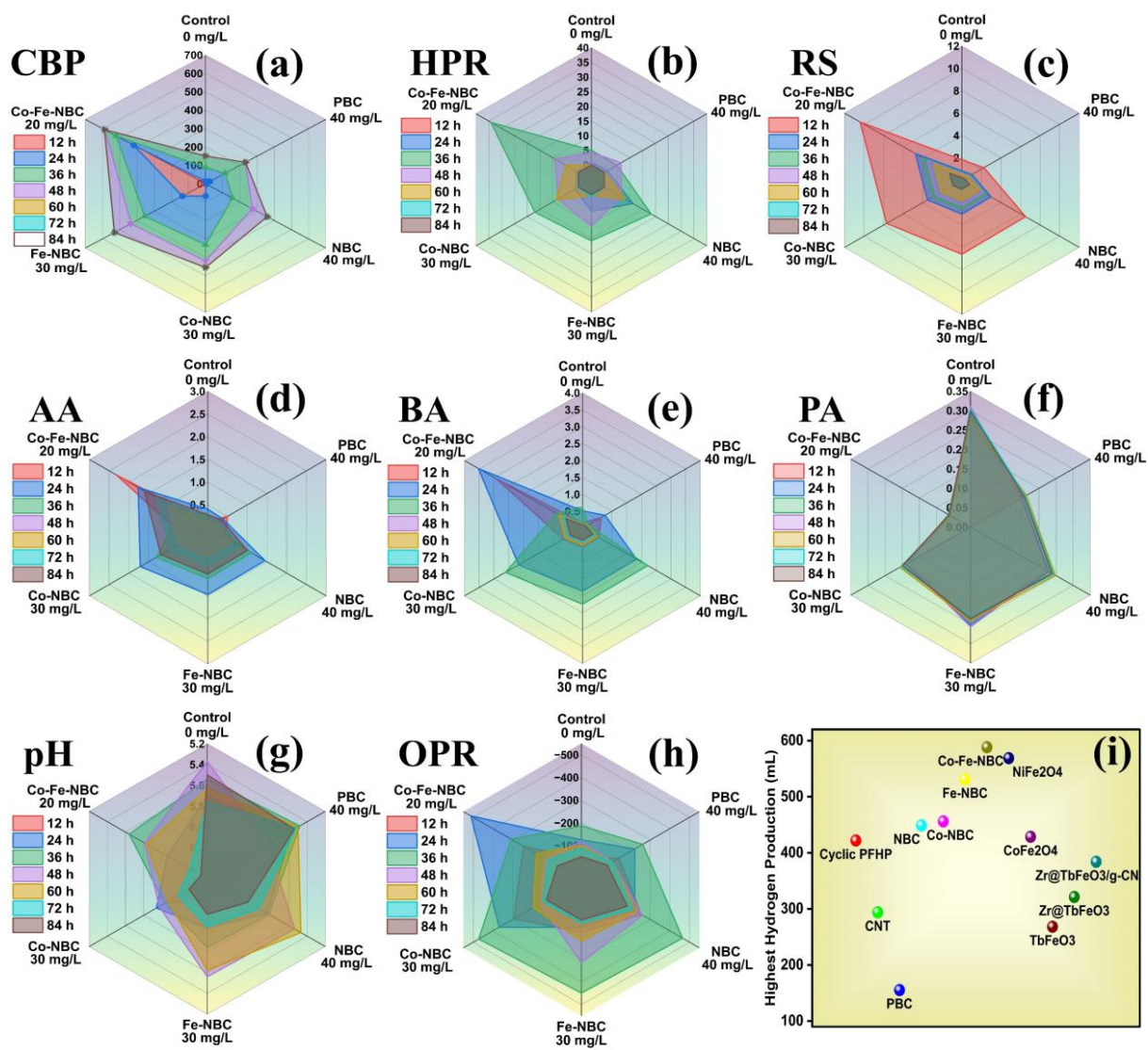

Fig. 5: Influence of NC on electron management on PFHP parameters (a) Cumulative Biohydrogen Production, (b) Hydrogen Production Rate, (c) Reducing Sugar, (d) Acetic Acid, (e) Butyric Acid, (f) Propionic Acid, (g) pH, (h) ORP, (i) Comparison of synthesized NC with previously published studies

### 3.4 Influence of biochar-based NC on dehydrogenase activity and NAD$^+$/NADH

Dehydrogenase activity constitutes a fundamental biochemical indicator of microbial electron transfer dynamics, operating through the oxidation of organic substrates coupled with the concomitant reduction of electron acceptors. Systematic supplementation of the fermentative medium with NC at their predetermined optimal concentrations elicited significant enhancements in dehydrogenase activity relative to the control system, as quantitatively demonstrated in Fig. 6(a). The control bioreactor maintained stable dehydrogenase activity throughout the experiment, consistently at 13.21 μg/mL, with a slight peak at 36 hours. In contrast, adding PBC led to a measurable increase, reaching a peak of 14.92 μg/mL. The NBC also caused a modest but clear rise to 15.87 μg/mL. Most importantly, bioreactors modified with transition metal-functionalized NC showed significantly higher dehydrogenase activity: Co-NBC reached 18.32 μg/mL, while Fe-NBC hit 21.69 μg/mL at 36 hours. The greatest catalytic boost appeared in the bimetallic Co-Fe-NBC system, achieving 24.73 μg/mL, an 87% increase over the control.

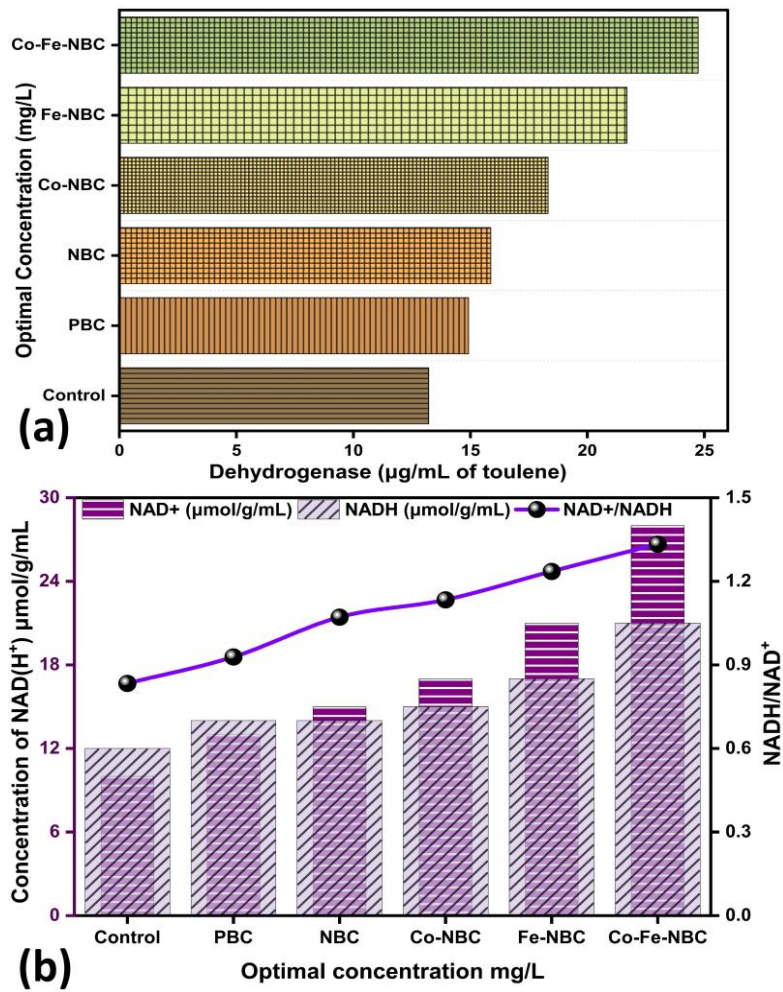

Fig. 6: Effect of NC on (a) dehydrogenase activity, (b) NAD$^+$ and NADH

This pronounced enzymatic potentiation is mechanistically attributable to augmented charge transfer efficiency within the fermentation matrix. The incorporation of redox-active metals (Co, Fe) within the Co-Fe-NBC architecture promotes the generation and mobility of free charge carriers [95], thereby facilitating electron shuttling during photo-fermentative hydrogen production (PFHP). Consequently, Co-Fe-NBC supplementation induced detectable modifications in dehydrogenase bio-electrochemical behavior and proton transfer kinetics, correlating directly with amplified metabolic flux. The elevated dehydrogenase activity accelerated core metabolic pathways, culminating in significantly enhanced biohydrogen yields (Fig. 5(a)) through two synergistic mechanisms: (i) dehydrogenase-mediated redox transformations that liberate $H^+$ ions via substrate interconversion, and (ii) fundamental restructuring of microbial consortia characteristics and functional composition [59, 96].

The analysis of pyridine nucleotide revealed parallel increases in the $NAD^+$/NADH system, essential cofactors controlling cellular redox homeostasis that act as key intracellular electron reservoirs for extracellular transfer processes (Fig. 6(b)) [97]. Control systems kept NADH concentrations steady at 10 µmol/g throughout the 24-36h operation period. The PBC and NBC caused moderate NADH buildup: PBC-based NC reached 13 µmol/g, while NBC-based NC hit 15 µmol/g. However, other metal-doped biochar caused significantly higher reducing equivalent storage: Co-NBC-based NC measured 17 µmol/g, Fe-NBC-based NC reached 21 µmol/g, and the bimetallic Co-Fe-NBC-based NC system showed exceptional NADH levels at 28 µmol/g. This rise in reduced cofactor availability directly links with the NADH-dependent hydrogenase pathway [98], a core mechanism for biohydrogen production. Additionally, these results indicate that NC-facilitated acceleration of $NAD^+$ regeneration kinetics boosts the release of intracellular electrons, creating thermodynamically favorable conditions for ongoing hydrogen generation [99]. Overall, the dehydrogenase activation and NADH increase phenomena form a clear biochemical framework explaining the markedly improved hydrogen production observed in systems amended with NC.

### 3.5 Enhanced electron management efficiency (EME) by biochar-based NC

Microbial biomass growth, VFAs generation, and a small amount of $H_2$ generation during hydrolysis are three types of electrons sinks in the PFHP system. From these three electron sinks, microbial mass growth and hydrolytic $H_2$ generation are considered minor electron sinks; therefore, it was assumed that these two were constant in all systems and could be omitted from the major EME calculation [6-8]. While VFAs' generation is a major electron sink, the whole EME calculations were based on VFAs concentrations of acetic acid (AA), butyric acid (BA), propionic acid (PA), and bioethanol (BE). From these VFAs, the sum of AA and BA is primarily considered favorable for biohydrogen production, while the sum of PA and BE is considered as an electron loss in the unfavorable biohydrogen production route [4, 14, 50]. For simplicity in EME calculations, the AA+BA is referred to as consumed electron, while PA+BE is referred to as diverted electron, as shown in Fig. 7(a, b, c), which depict the number of diverted electrons, consumed electrons, and the highest electron management efficiency achieved, respectively. At the same time, detailed calculations and results related to EME are elaborated in Table S (1-4), which exhibited overall performance: control (8.9% EME, 0.01348 mol e⁻ used 0.1378 mol e⁻ wasted), PBC (23.4% EME, 0.02080 mol e⁻ used, 0.0679 mol e⁻ wasted), NBC (38.3% EME, 0.03206 mol e⁻ used, 0.0517 mol e⁻ wasted), Co-NBC (49.8% EME, 0.04072 mol e⁻ used, 0.0411 mol e⁻ wasted), and Fe-NBC (63.3%, 04740 mol e⁻ used, 0.0274 mol e⁻ wasted).

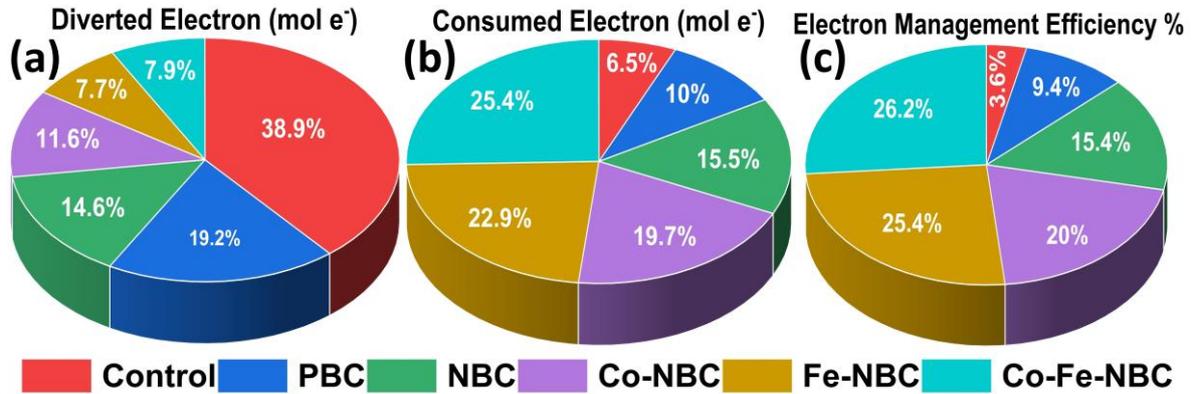

Fig. 7: Effect of biochar-based NC on (a) diverted electrons, (b) consumed electrons, and (c), electron management efficiency

The control shows the lowest EME as the PFHP system follows the natural flow system. In contrast, PBC, NBC, Co-NBC, Fe-NBC and Co-Fe-NBC show higher EME due to their higher physicochemical and electrochemical properties to handle electron flux. The Co-Fe-NBC shows an excellent EME in PFHP, directing 65.3% of typical electrons toward biohydrogen production (05264 mol e⁻ used for H$_2$) while minimizing waste to competing pathways (0.0280 mol e⁻ wasted), a 7.3-fold improvement over the control. The superior electron channeling in Co-Fe-NBC is attributed to its hybrid capacitive-redox mechanism: capacitive storage (287 F/g) buffers electron overflow during metabolic surges, while $Fe^{2+}/Fe^{3+}$ and $Co^{2+}/Co^{3+}$ redox cycles (15.8 mC/g capacity) act as dynamic reservoirs, releasing stored electrons during shortages. This dual functionality minimized electron leakage to propionate (PA: 0.31 g/L, 0.0084 mol e⁻ wasted) and bioethanol (BE: 0.45 g/L, 0.0196 mol e⁻ wasted), lowering overall waste by 80% compared to the control (PA: 2.07 g/L, 0.0558 mol e⁻; BE: 1.89 g/L, 0.0820 mol e⁻). In comparison, Fe-NBC, despite its high efficiency (63.3%), produced more PA waste (0.45 g/L, 0.0122 mol e⁻) than Co-Fe-NBC, highlighting the essential role of iron-cobalt-nitrogen synergy in stabilizing redox intermediates. The nitrogen-doped carbon matrix of Co-Fe-NBC further improved direct electron transfer (Rct 6 Ω), bypassing soluble mediators and maintaining continuous electron flow to nitrogenase. This precise mechanistic function of Co-Fe-NBC confirms its capattery-like design as the most effective electron control system. Evaluation showed that NBC had poor electron shuttling efficiency, at 23-38%, due to a lack of redox-active sites. The results clearly demonstrate that Co-Fe-NBC's multifunctionality overcomes PFHP's electron leakage issue, reaching industrial-grade electron efficiency (> 65%) through controlled storage, redox buffering, and targeted electron delivery. Meanwhile, single-doped or pristine biochar lacks the synergistic functions needed for effective flux regulation.

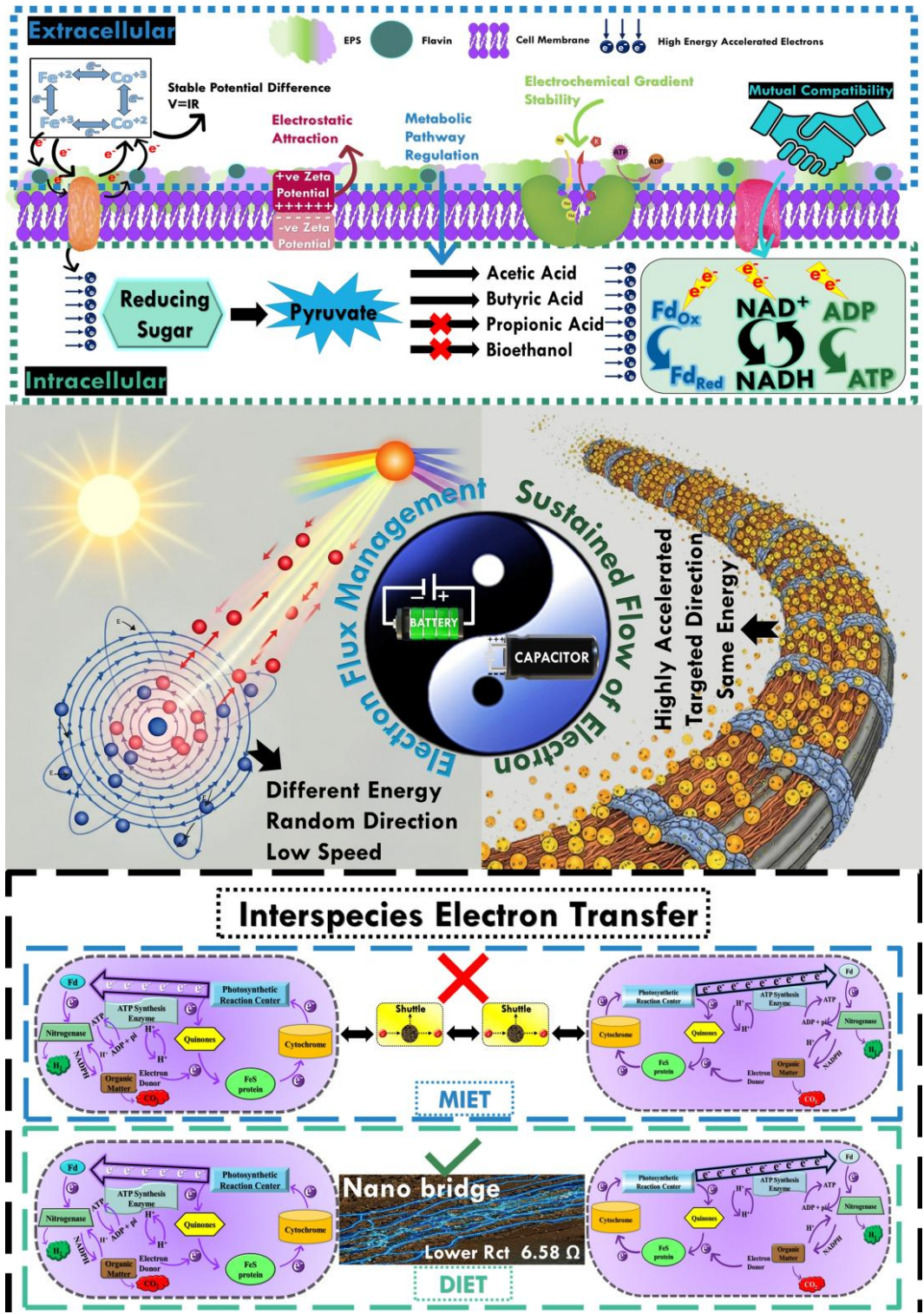

Fig. 8: Mechanistic overview of NC interactions in PFHP system

The transformative performance of biochar-based NC (Co-Fe-NBC) is attributed to its sophisticated resolution of the fundamental electron management issues in PFHP. It establishes a dynamic system that interfaces directly with microbial metabolism by creating a conductive nitrogen-doped matrix that acts as a low-resistance (6.58 Ω) electron highway (Fig. 8). This direct pathway allows electrons to flow efficiently from microbial cytochromes to the catalyst surface, effectively bypassing slower soluble electron shuttles. The material's true innovation lies in its nano-capattery characteristic, synergizing rapid capacitive charge storage with sustained Faradaic discharge from $Fe^{2+}/Fe^{3+}$ and $Co^{2+}/Co^{3+}$ redox cycles. This dual functionality provides critical electron buffering during metabolic peaks and a steady release in later stages. Such precise electron-level control stabilizes the intracellular redox environment and elevates the $NAD^+/NADH$ ratio (1.34), which actively prevents metabolic shifts toward propionic acid production. Furthermore, the hierarchical porosity supports robust microbial colonization while its tailored surface chemistry offers localized pH buffering to maintain optimal activity. The synergistic effect of these mechanisms culminates in a profound metabolic reprogramming, efficiently capturing and channeling reducing equivalents toward biohydrogen production. Consequently, Co-Fe-NBC functions as an intelligent electron management system that actively interacts with the microbial consortium across electronic, enzymatic, and metabolic levels. The detailed discussion of results is presented in the supplementary data, and a mechanistic overview of Co-Fe-NBC-based NC for electron management and metabolic pathway regulation is presented in Fig. 8.

**Conclusion**

This study shows that the Co-Fe-NBC-based NC is an effective tool for controlling electrons to improve biohydrogen production. It functions both as a battery and supercapacitor, storing excess electrons and releasing them gradually and controlled. This dual role stabilizes the entire PFHP system and prevents unwanted by-products by managing electron flow. More than 65% of electrons are directed towards hydrogen production, reprogramming microbial metabolism to favor beneficial acids like acetate and butyrate, which are linked to high hydrogen yields, while reducing competing pathways by about 85%. Its porous structure creates an ideal environment for microbial growth, increasing activity and maintaining stability. As a result, hydrogen production rises by approximately 367%. This research offers a practical, efficient approach for electron management in complex biological systems, supporting scalable, clean energy solutions to help lower our carbon footprint.

**Author contributions**

Muhammad Shahzaib: Writing-original draft, Software, Formal analysis, Data curation, Methodology. Faiqa Nadeem: Review & editing. Muhammad Usman: Review & editing. Muneeb Ur Rahman: Review & editing. Hina Ramzan: Review & editing. Nadeem Tahir: Writing-review & editing, Supervision, Project administration, Funding acquisition, Conceptualization.

**Acknowledgment**

The authors gratefully acknowledge the financial support of the Henan Provincial Joint Fund (Applied Research Category (242103810015)).

**Data availability**

No data were used in the research from other studies described in the article. Data of this study will be provided on demand.

**Declaration of Competing Interest**

The authors declare that they have no known competing financial interests or personal relationships that could have appeared to influence the work reported in this paper.